\documentclass[10pt,twocolumn]{article} 
\usepackage{simpleConference}
\usepackage{times}
\usepackage{graphicx}
\usepackage{amssymb}
\usepackage{url,hyperref}
\usepackage{siunitx}
\usepackage{tikz}
\usepackage{epstopdf}

\begin{document}

\title{\huge High-speed extended-volume blood flow measurement using engineered point-spread function}

\author{Yongzhuang Zhou, Vytautas Zickus, Paul Zammit, Jonathan M. Taylor, and Andrew R. Harvey{*} \\
School of Physics \& Astronomy, University of Glasgow, Glasgow, UK G12 8QQ \\
\\
andy.harvey@glasgow.ac.uk  \\
}

\maketitle
\thispagestyle{empty}

\begin{abstract}
Experimental characterization of blood flow in living organisms is crucial for understanding the development and function of cardiovascular systems, but there have been no techniques reported for snapshot imaging of thick samples in large volumes with high precision. We have combined computational microscopy and the diffraction-free, self-bending property of Airy beams to track fluorescent beads with sub-micron precision through an extended axial range (up to \SI{600}{\micro\meter}) within the flowing blood of 3 days post-fertilization (dpf) zebrafish embryos. The spatial trajectories of the tracer beads within flowing blood were recorded during transit through both cardinal and intersegmental vessels, and the trajectories were found to be consistent with the segmentation of the vasculature recorded using selective-plane illumination microscopy (SPIM). This method provides precise spatial and temporal measurement of 3D blood flow with potential for enhanced understanding of fundamental dynamics, such as measurement of wall shear stress or of cardiovascular disease.
\end{abstract}

\section{Introduction}
Measurement of the spatio-temporal properties of blood flow in the cardiovascular system is crucial for understanding of cardiac morphogenesis~\cite{Hove2003,Jamison2013,Forouhar:06},
angiogenesis and vasculogenesis~\cite{Ramasamy2016,Korzh2008,Watkins2012,ROSHTHKARI:11}, since the early vascular formation is believed to be not only genetically predetermined but also governed by external mechanical stimuli. Recent studies have also revealed that blood flow is a key factor for controlling aging processes in the skeletal system~\cite{Ramasamy2016}, and plays an important role in brain functioning~\cite{Mintun2001,Dunn2001,Fujishima1995} and in the continued growth of  organs such as the liver~\cite{Korzh2008}. Additionally, flow-induced forces, such as wall shear stress and transmural pressure, are believed to have an important influence on heart development and valve formation~\cite{Hove2003,Jamison2013}. Research into cardiovascular dynamics is often focused on the zebrafish embryo due to its genetic relevance, small size and transparency~\cite{Lieschke2007,Kari2007}, but to date there have been no techniques reported that are able to record 3D blood flow \textit{in vivo} throughout extended volumes of  zebrafish embryos with sufficient time resolution to directly resolve  typical pulsatile hemodynamics.

A wide range of techniques have previously been reported for measuring aspects of blood-flow dynamics, but they suffer from limitations.  Fluorescence correlation spectroscopy (FCS) employs confocal laser scanning \cite{Pan2007,Pan2009,Shi2009} to deduce blood velocities from the temporal intensity fluctuations of fluorescence and the scanning direction of the laser focus. FCS can provide relatively high spatial resolution but is restricted  to low concentrations and small observation volumes~\cite{Fieramonti2015}, and the point-scanning nature of the imaging makes it unsuitable for time-resolved imaging of, for example, the pulsatile flow in the cardiovascular system.
Similarly, optical vector field tomography (OVFT), which combines optical projection tomography (OPT) with high-speed multi-view acquisition and particle image velocimetry (PIV)~\cite{Fieramonti2015}, can produce a 3D velocity map of  blood flow at the whole organism level, but the requirement to rotate the sample during data acquisition prevents high-speed operation. 
We have recently  demonstrated selective-plane-illumination microscopy in conjunction with micro PIV (SPIM-$\mu$PIV) and high-speed, heart-synchronised,  multi-depth acquisition to enable 3D measurement of blood flow in zebrafish~\cite{Zickus:18}, but the plane-by-plane optically-sectioned acquisition precludes easy measurement of the axial component (third dimension) of the flow vectors.

\begin{figure*}[ht]
\centering
\includegraphics[width=0.13\linewidth]{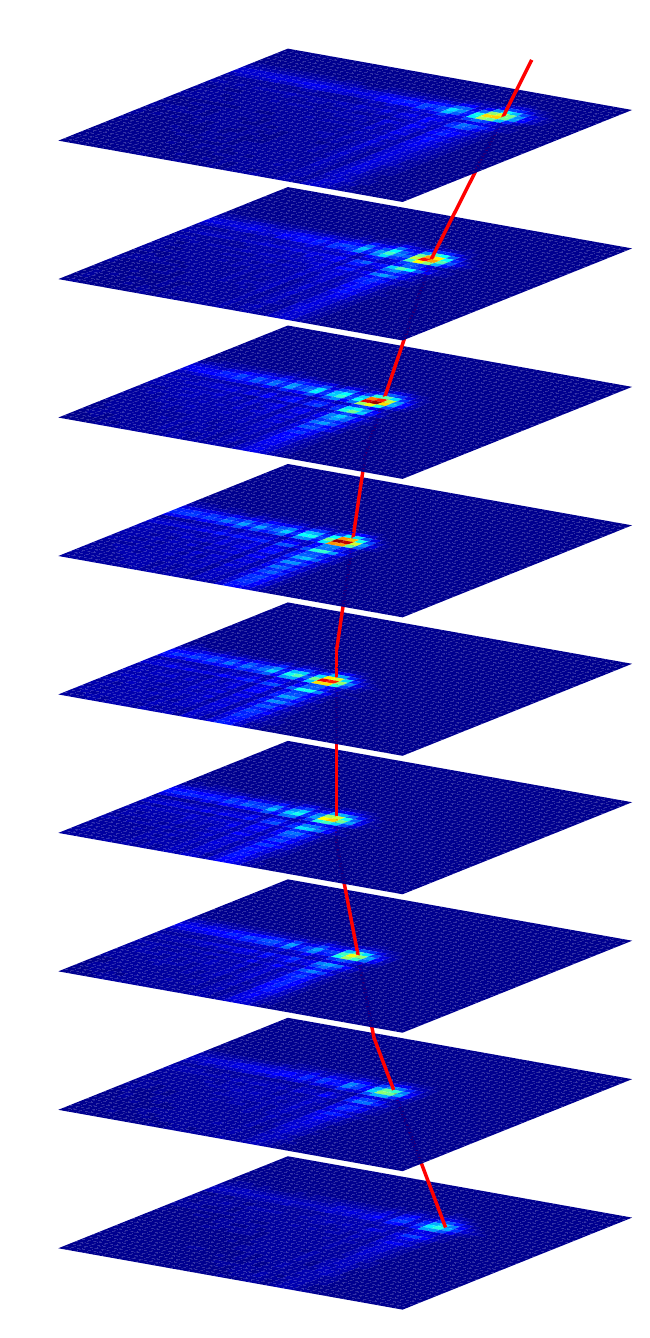}\ \ \ 
\includegraphics[width=0.13\linewidth]{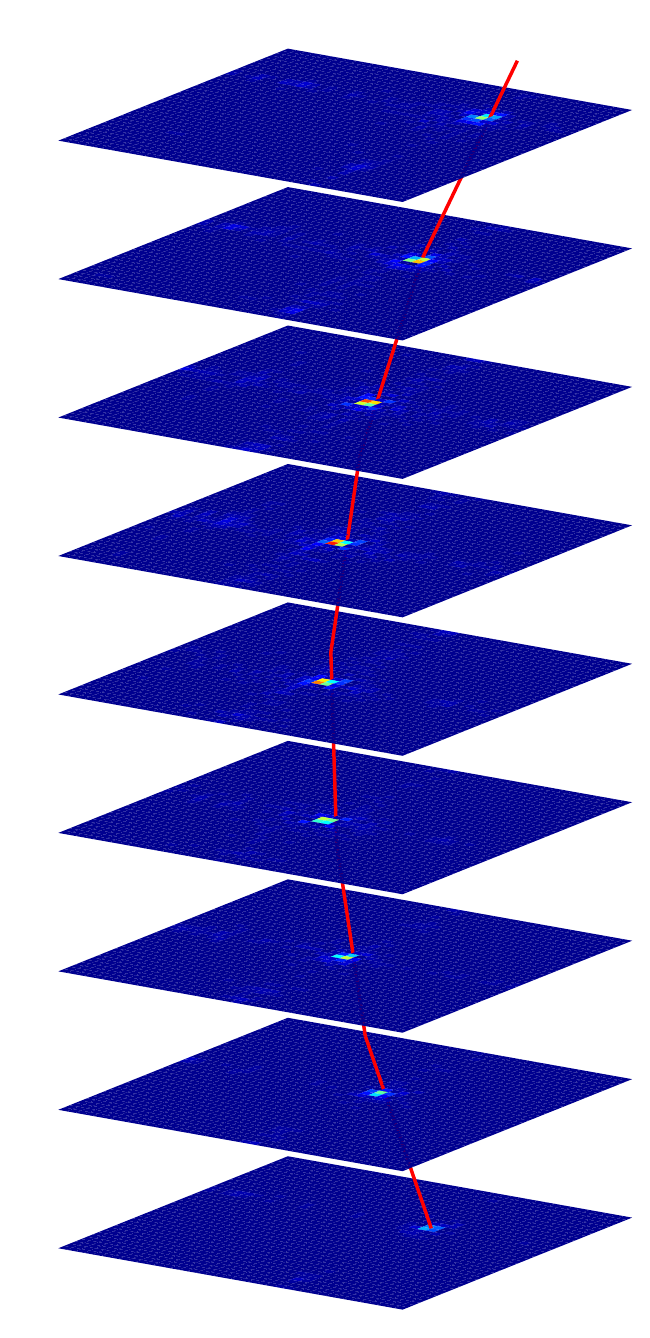}\ \ \ \ \ \ \ 
\includegraphics[width=0.46\linewidth]{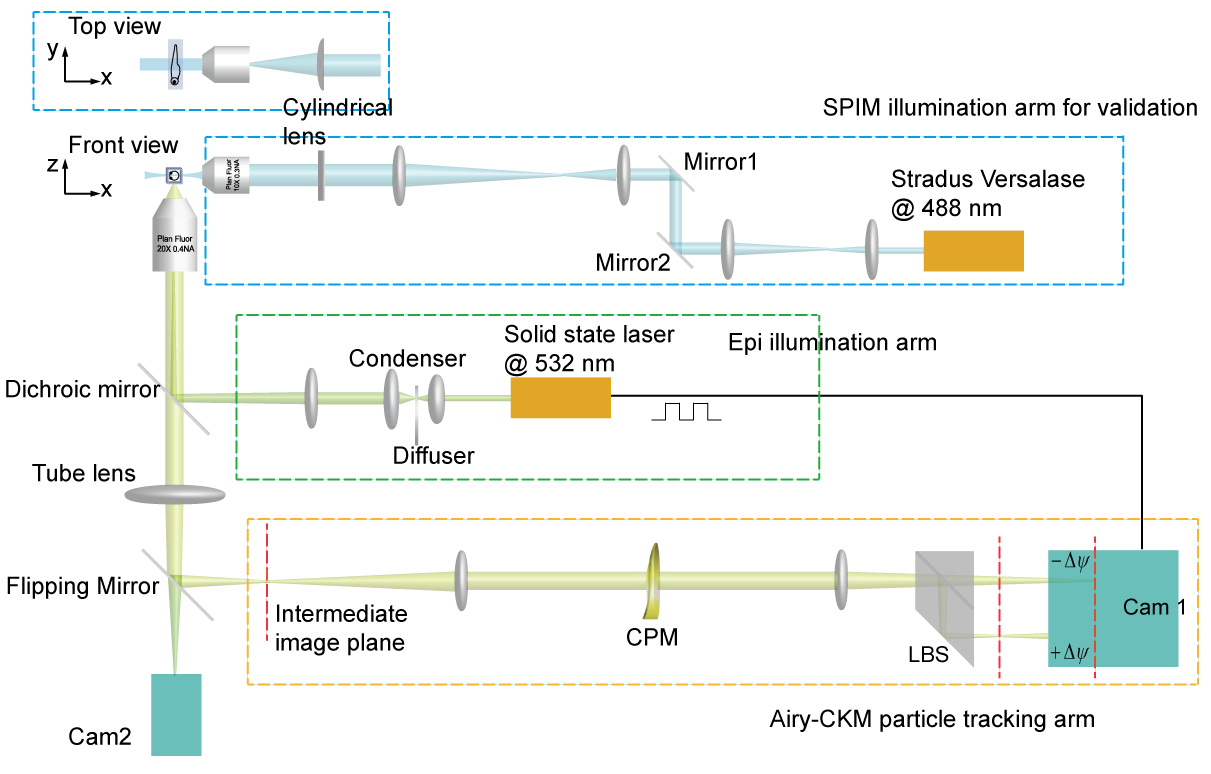}
\put(-350,-12){(a)} \put(-280,-12){(b)} \put(-120,-12){(c)}
\caption{\small (a) Airy-beam PSFs generated with an $\alpha = 7$ cubic phase mask over a depth range of \SI{200}{\micro\metre}, with the red curve indicating the accelerating translation along the image diagonal. (b) Corresponding recovered PSFs using Wiener deconvolution with the in-focus PSF used as the deconvolution kernel. (c) Schematic of the experimental setup. The dashed orange box indicates the the particle tracking arm.  A 0.5 NA, 20$\times$ CFI Plan Fluor objective was used to image the tracer particles. A cubic phase mask was placed at the re-imaged pupil plane of a 4-$f$ relay system. The lateral beam splitter then generates two images of the same scene on the same camera sensor (camera 1), but with two opposite defocus offsets. Epi illumination was used for the 3D particle tracking, with a 532 nm solid state laser as the light source. The dashed blue box shows the SPIM illumination arm used for validation imaging (acquired with camera 2). The light sheet was launched with a 10$\times$, 0.3 NA CFI Plan Fluor objective and a 75 mm cylindrical lens. The sample was mounted in a square glass capillary on a piezo $z$ stage. CPM: cubic phase mask; LBS: lateral beam splitter.}
\label{exp_setup}
\end{figure*}

On larger length scales, Doppler optical coherent tomography (DOCT) has been widely used to characterize human retinal hemodynamics~\cite{LEITGEB2014}. Although preliminary work has also been reported for visualizing blood flow in zebrafish~\cite{Iftimia2008}, the theoretical resolution is at best several \SI{}{\micro\meter}, and the effective resolution is severely limited by speckle noise, requiring integration over a significant portion of the cardiac cycle. Furthermore, DOCT is only able to measure one velocity component (parallel to the probe beam). This frustrates attempts at quantitative flow measurement, which require \emph{a priori} structural information, or complex multi-beam configurations in order to measure the full vector velocity field~\cite{Trasischker2013} at the cost of transverse resolution. 

Lu et al. took a different approach to real-time 3D imaging using defocusing digital particle imaging velocimetry (DDPIV) for \textit{in vivo} blood-flow characterization~\cite{Lu2010} using  microinjected tracer particles. DDPIV employs a three-pinhole mask at the pupil plane to optically encode 3D particle position of tracer particles as in the form of 2D images on a detector array \cite{Pereira2002}. Such a three-pinhole mask, however, severely limits the numerical aperture and optical throughput of the imaging system, yielding a reduced signal-to-noise ratio (SNR) and localization precision. Moreover, the rapid expansion of the PSF with defocus severely restricts the maximum seeding concentration and axial range (about \SI{40}{\micro\meter} as reported).

A potential solution to these limitations lies in the use of pupil-engineered localization microscopy, which can provide localization of point emitters with a precision of tens of nm and has been widely used in super-resolution microscopy, single-particle tracking in cells and microfluidics experiments~\cite{Pavani:09,Jia:14,Zhou:18}. 

We report the first application of pupil-engineered localization microscopy to \textit{in vivo} blood-flow characterization; in particular we use the Airy-CKM technique~\cite{Zhou:18} to map blood flow within 3-dpf zebrafish with sub-100 nm precision and sub-1 ms temporal resolution. The inherent diffraction-free range of the Airy-beam PSF provides an extended detection volume that encompasses the whole thickness of the zebrafish body (up to \SI{400}{\micro\meter}$\times$\SI{700}{\micro\meter}$\times$\SI{600}{\micro\meter} when using a cubic-phase mask with phase parameter $\alpha=12$, i.e. the peak modulation introduced by the phase mask in waves). Computational recovery of a compact PSF from the extended Airy-beam PSF allows a higher seeding concentration to be achieved than is possible with other optical-encoding techniques. Trajectories and velocities of fluorescent tracer beads within blood in both the cardinal and intersegmental vessels are observed, and the reconstructed trajectories are validated by quasi-simulataneous 3D sectioning of the zebrafish vascaulature using a modified SPIM system~\cite{Greger2007}.

\section*{Results}

\subsection*{3D Mapping of \textit{in vivo} blood flow with Airy-beam PSFs} 

Our Airy-CKM technique exploits four optical properties of the Airy-beam PSF: (1) the parabolic transverse translation of the Airy-beam PSF  with defocus, (2) the extended range of diffraction-free propagation  (3) the absence of zeros in the modulation-transfer function (MTF)\cite{Dowski:95,Zammit:14,Demenikov:10}. This enables robust recovery  of compact, diffraction-limited PSFs that encode axial displacement over an extended axial range. Fig. 1(a) shows a stack of PSFs acquired near the nominal focal plane, with defocus ranging from \SI{-100}{\micro\metre} to  \SI{100}{\micro\metre}. The red curve indicates the lateral translations of the PSF; in the Airy-CKM technique, this image translation is exploited to calculate the depth coordinate of each point emitter~\cite{Zhou:18}. The shape of the PSF is approximately propagation-invariant, thus allowing Wiener deconvolution to be performed with a single kernel, yielding the compact profile shown in Fig. 1(b). A PSF deconvolved in this way exhibits the same lateral translation as the Airy-beam PSF, but with a higher peak signal level and a compact intensity profile that approximates a diffraction-limited PSF. 

In our experimental setup, the Airy-beam PSF is obtained by placing a cubic phase mask either within the pupil plane of the microscope or at a re-imaged pupil plane as shown in Fig. 1(c). The depth-dependent image translation is measured by duplication of the image with different defocus offsets on the same camera sensor using a lateral beam splitter (LBS), and particle coordinates are determined from the relative displacement of the two images of each particle as discussed in the Methods section. 

\begin{figure}[ht!]
\centering
\includegraphics[width=0.45\linewidth]{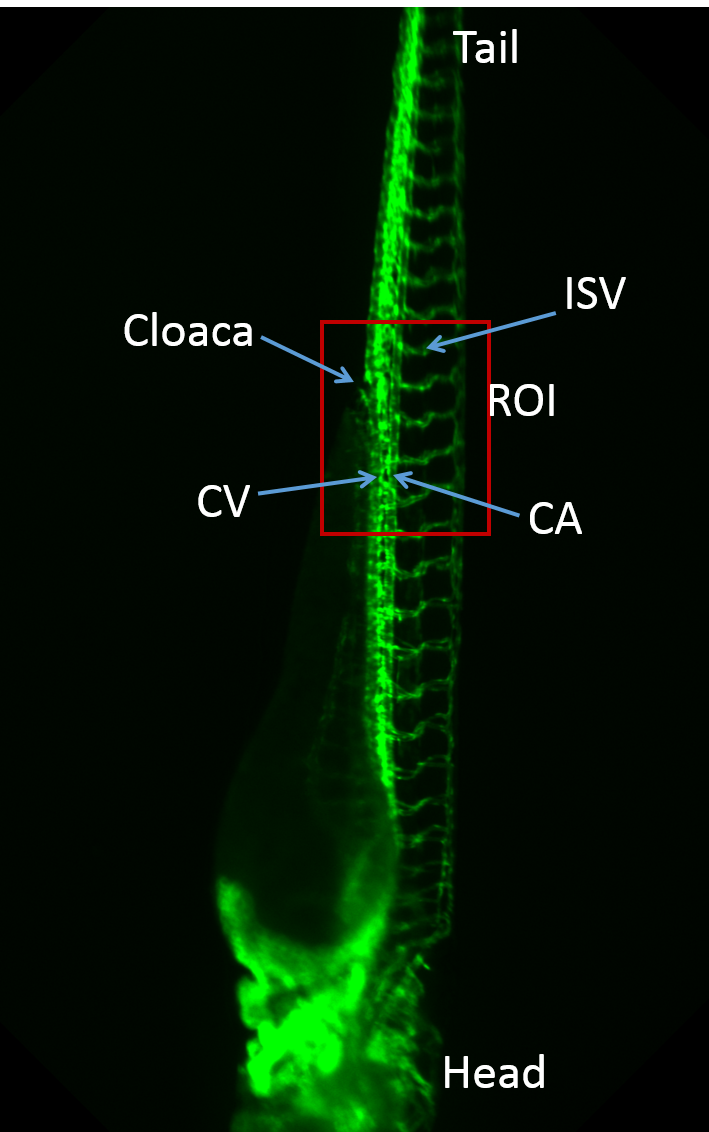}\ \ 
\includegraphics[width=0.462\linewidth]{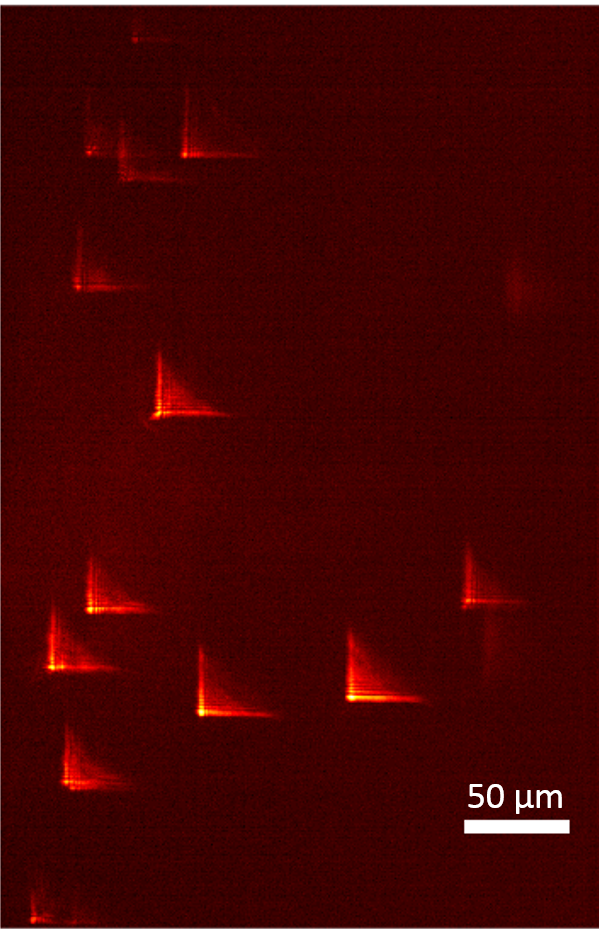}\\  
\includegraphics[width=0.85\linewidth]{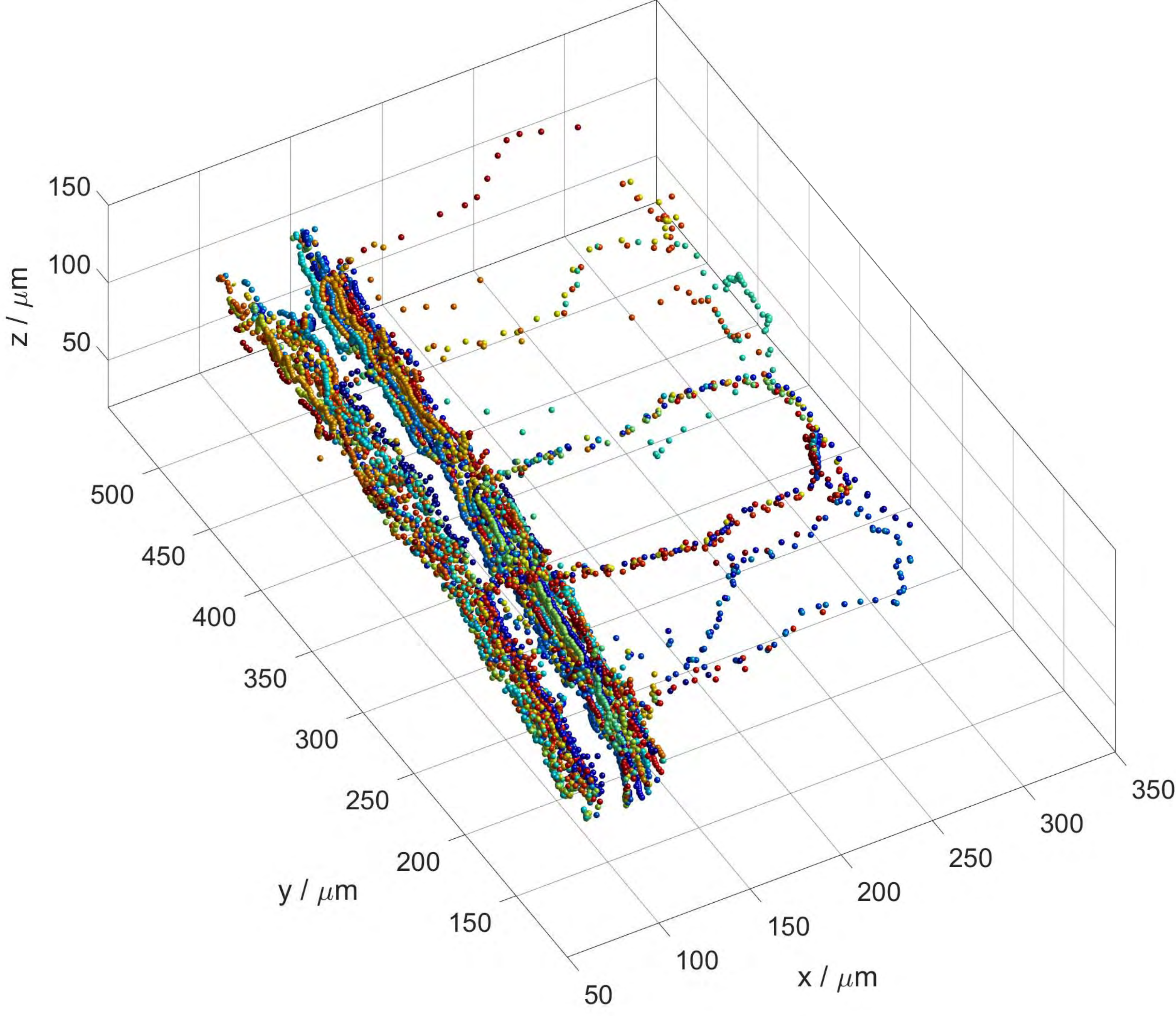}
\put(-160,170){(a)} \put(-60,170){(b)} \put(-90,-12){(c)}
\caption{\small (a) Epifluorescence image of a flk1:GFP zebrafish embryo, showing the region of interest (ROI) used in our experiment. CV: cardinal vein, CA: cardinal artery, ISV: intersegmental vessel. (b) PSF encoded image showing 1 µm fluorescent tracer beads flowing within the ROI. (c) 3D trajectories of tracers within cardinal and intersegmental vessels of zebrafish, reconstructed from 2000 frames. Color-coding is used to distinguish the trajectories of each tracer particle. Supplementary \textcolor{blue}{Visualization 1} shows the tracers flowing with the blood in real-time, along with a 3D reconstruction of tracer trajectories.}
\end{figure}

We performed \textit{in vivo} 3D blood-flow imaging of \SI{1}{\micro\metre}-diameter fluorescent beads micro-injected as tracers within both the cardinal vessels and the intersegmental vessels of 3 dpf zebrafish. The tracer trajectories were reconstructed to yield the spatial and temporal 3D-velocity vectors of the blood flow within the zebrafish vasculature. Figure 2(a) shows the blood vessels of a 3 dpf zebrafish recorded with a stereo microscope in green fluorescence, with the quasi-periodic structure of the intersegmental vessels visible along the length of the tail. The selected region of interest (ROI) was near the cloaca as indicated by the red rectangle in Fig. 2(a). This encompasses several intersegmental vessels, the cardinal artery and the cardinal vein which are of interest for this investigation.

The injected fluorescent beads flow with the blood and appear as high-contrast PSFs as shown in Fig. 2(b), which depicts a still image of the flowing tracers taken from a 2000-frame video sequence (Dataset 1\cite{ZhouData2018}). Each Airy-beam PSF is the image of a single bead in the bloodstream. The 3D tracer locations in each frame can be extracted from the Airy-beam PSFs using the algorithm discussed in the Methods section. The field of view is about \SI{400}{\micro\metre} $\times$ \SI{700}{\micro\metre}, and the extended depth range provided by the Airy-beam PSF allows the blood flow throughout the whole thickness of the fish (normally more than \SI{100}{\micro\metre}) to be recorded in a single snapshot without optical sectioning. This permits a high volume-imaging rate: up to hundreds of snapshot volumes per second depending on the ROI, and limited only by the camera performance. 

We recorded video sequences of the blood flow at a frame rate of 26.5 frames per second (fps), and tracked the trajectory of each bead within the image sequence using the algorithm developed by Crocker and Grier~\cite{Crocker1996}. The displacements between two successive frames were used to approximate the instantaneous velocities. Figure 2(c) shows the reconstructed 3D trajectories of the tracers flowing within intersegmental and cardinal vessels in the ROI, where the coloring indicates trajectories of individual beads. As shown, most tracers flow through the cardinal artery and cardinal vein with a few tracers entering the intersegmental vessels supplying the surrounding muscle tissue. The 3D structures of the intersegmental vessels and capillaries can be deduced from the trajectories of the injected tracers flowing through them.

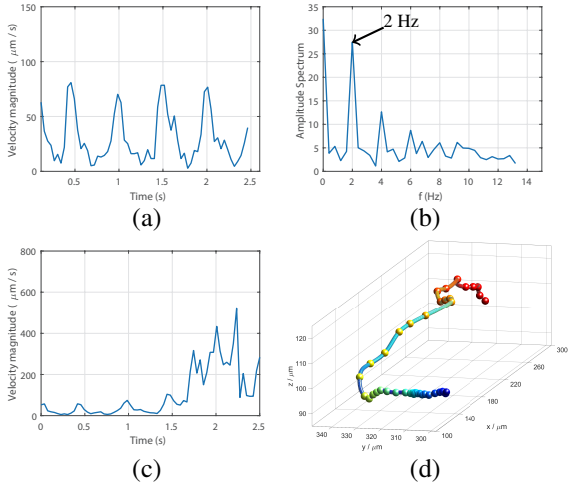
\begin{figure}[ht]
\centering
\begin{tikzpicture}
\node[anchor=south west,inner sep=0] at (0,0) {\includegraphics[width=0.45\linewidth]{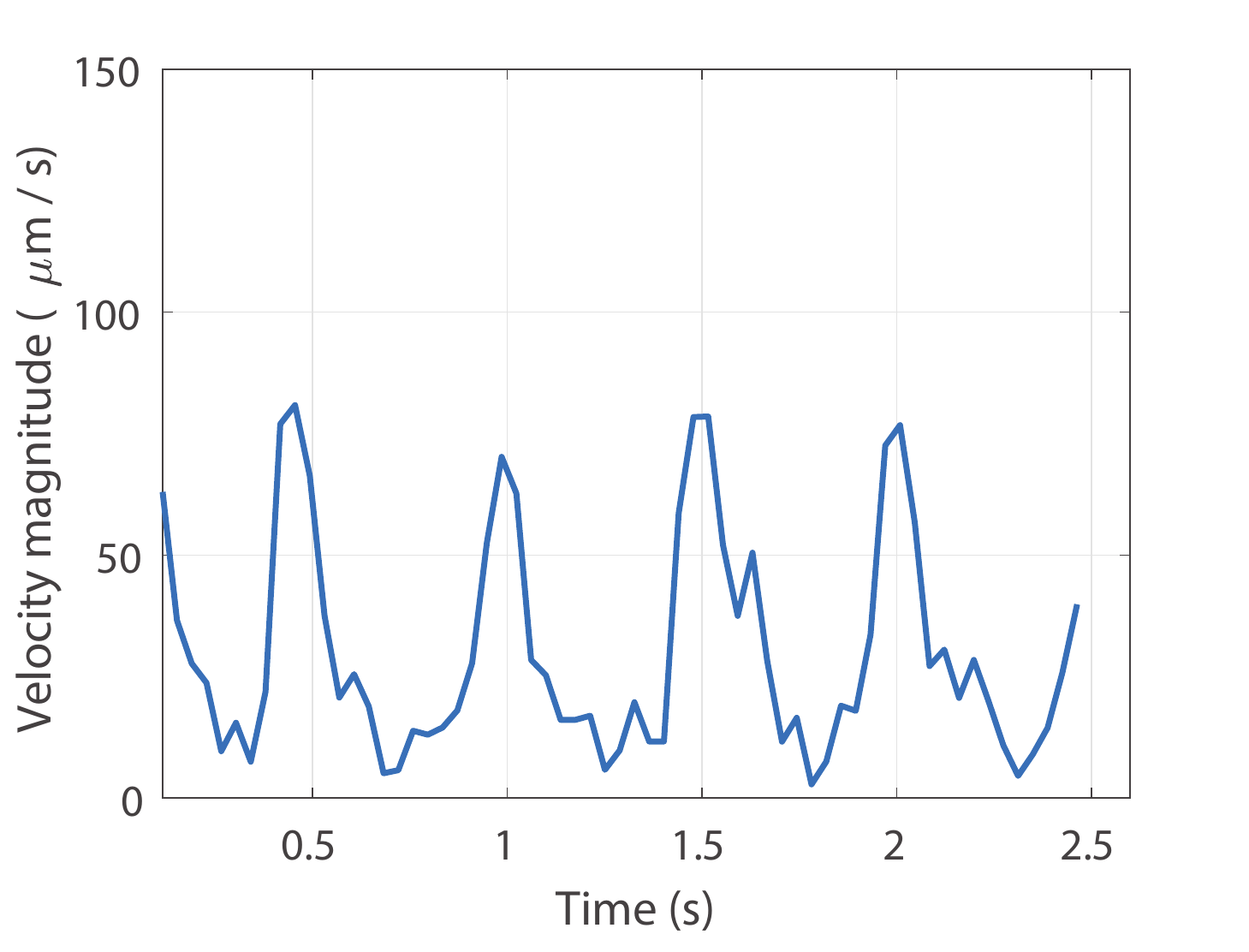}\includegraphics[width=0.45\linewidth]{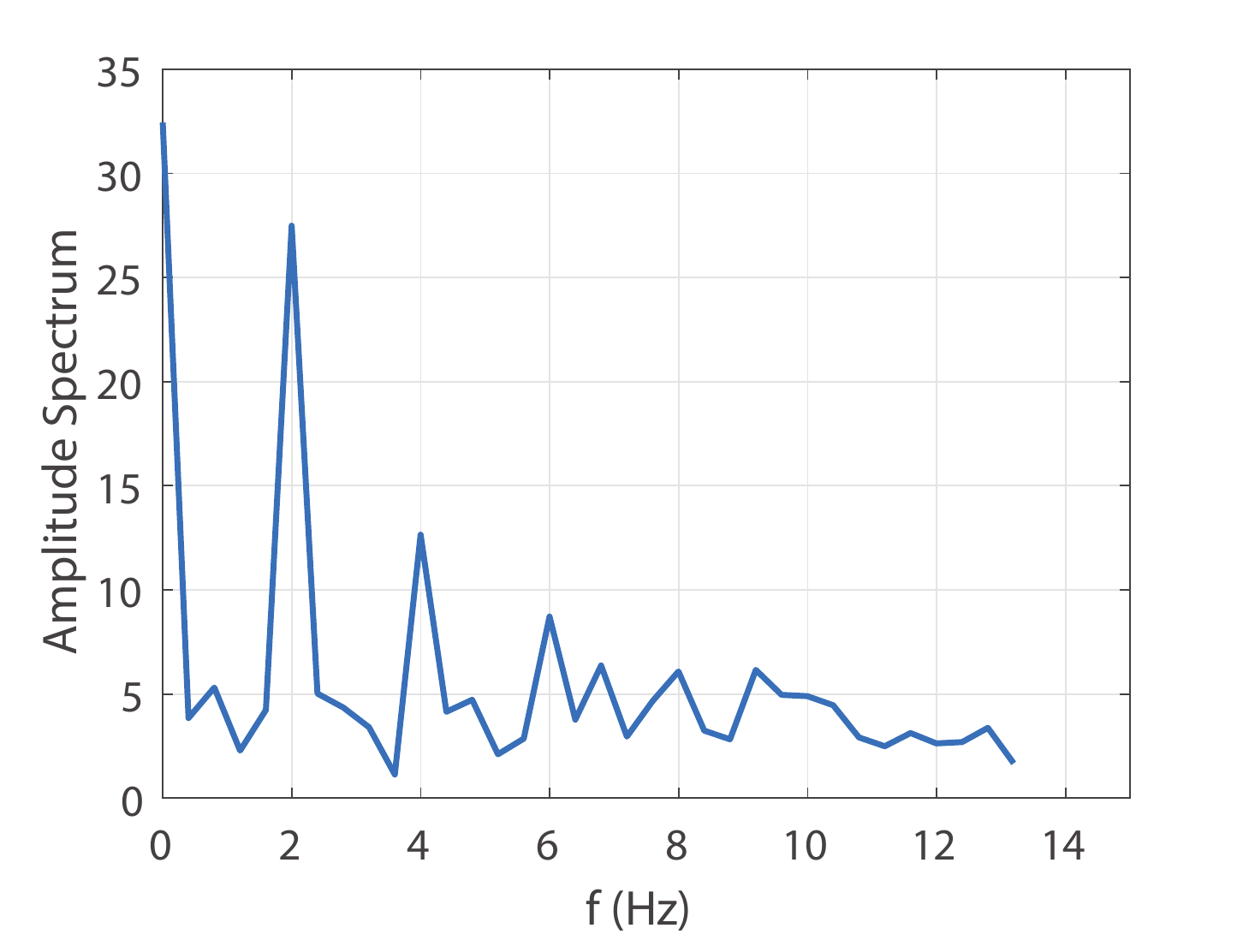}};
\node[anchor=south west,inner sep=0] at (0,-3.25) {\includegraphics[width=0.45\linewidth]{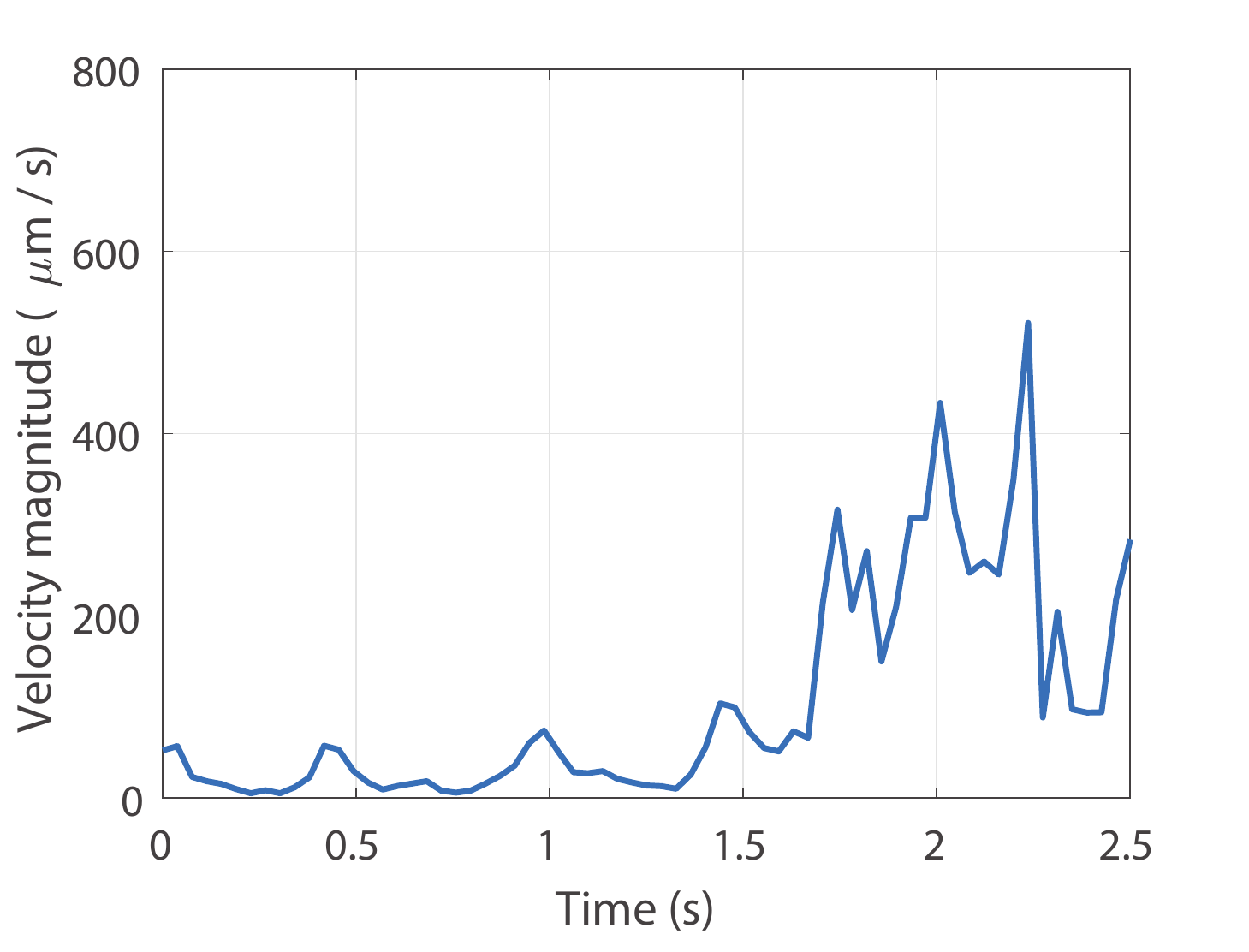}\includegraphics[width=0.45\linewidth]{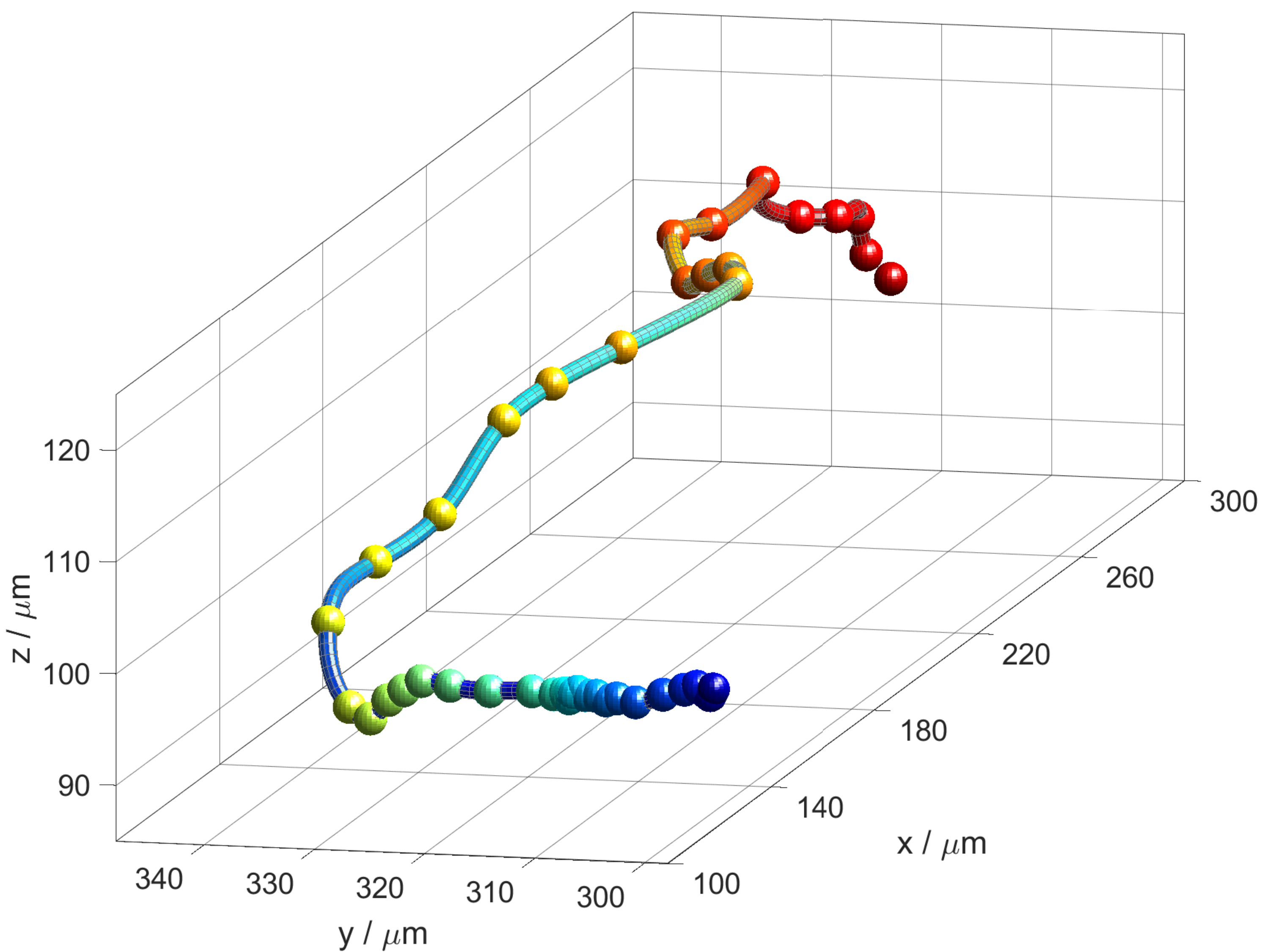}};
\draw[thick,->] (5.05,2.38) -- (4.63,2.18) node[anchor=south east] at (5.65,2.25) {\scriptsize 2 Hz};
\end{tikzpicture}
\put(-165,85){(a)} \put(-60,85){(b)} \put(-165,-10){(c)} \put(-60,-10){(d)}
\caption{\small Temporal and spatial variation of the tracer velocity. (a) Velocity of a tracer flowing within the cardinal artery. (b) Fourier spectrum of the velocity profile in (a). (c) Velocity of a tracer that enters an intersegmental vessel from the cardinal vessel. (d) 3D trajectory of the tracer in (c), color-coded for depth.}
\label{fig: graphs}
\end{figure}

Figure 3(a) shows the velocity magnitude of a single bead flowing within a cardinal artery as a function of time, and Fig. 3(b) shows its associated Fourier spectrum. The velocity variation of the tracer can be seen to be quasi-periodic, with a fundamental frequency of about 2 Hz corresponding to the heart rate of the zebrafish.
Since the injected tracers are relatively light and small, they have negligible influence on the function of the zebrafish heart, and monitoring of the heart rate also indicated unperturbed conditions throughout our experiment.
Figure 3(c) shows the change in velocity of a tracer particle as it enters an intersegmental vessel from a cardinal vessel, of which the trajectory is shown in Fig. 3(d). As can be observed, the peak velocity increased dramatically from below \SI{100}{\micro\metre.s^{-1}}  to about \SI{600}{\micro\metre.s^{-1}} due to the narrowing of the vessel diameter.

\begin{figure}[ht!]
\centering
\includegraphics[width=0.6\linewidth,trim={7cm 4.5cm 4cm 3cm},clip]{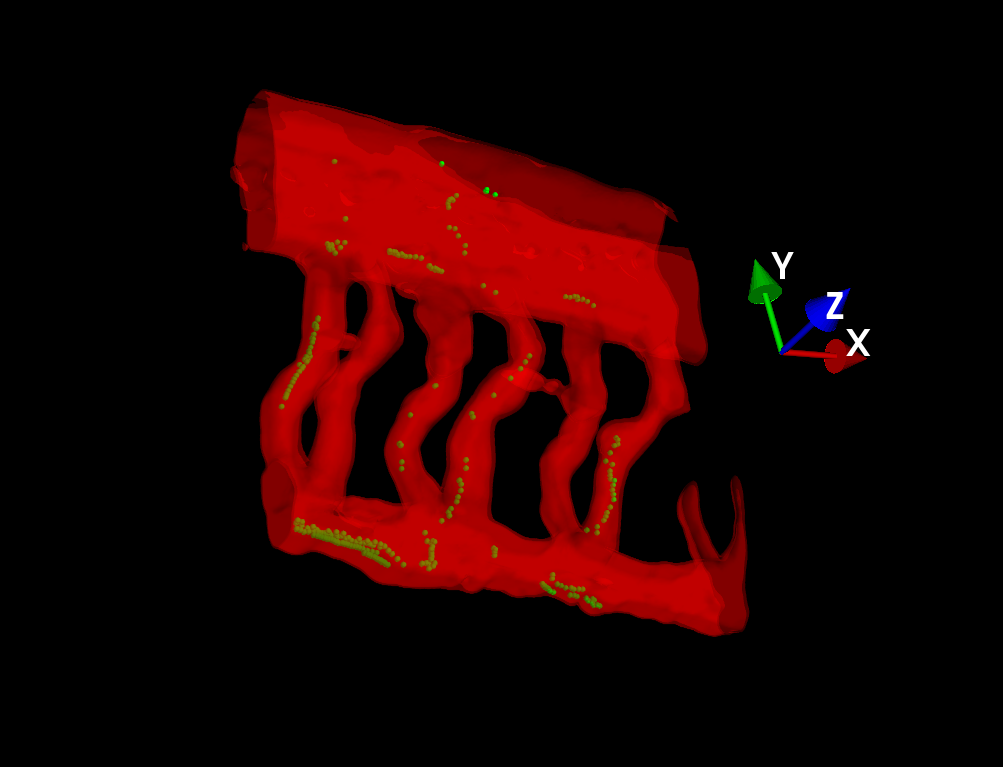}\ \includegraphics[width=0.39\linewidth,trim={3cm 0cm 1cm 0cm},clip]{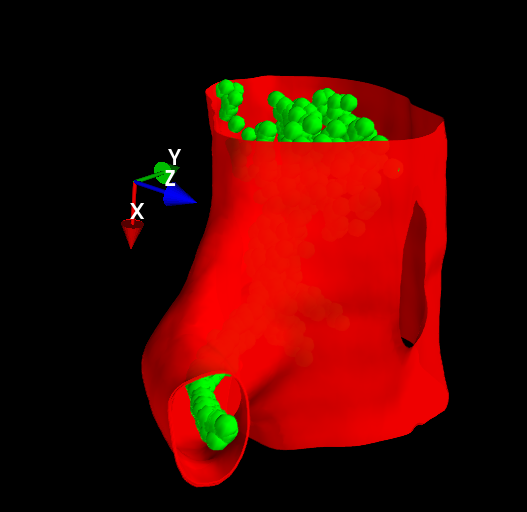}
\put(-180,-10){(a)} \put(-60,-10){(b)}
\caption{\small Composite images showing detected tracer locations (green spheres) embedded within a surface rendering (red) of the 3D blood vessel structure obtained from quasi-simultaneous SPIM imaging (Dataset 1\cite{ZhouData2018}). (a) View of several tail segments; (b) detail of a vessel junction. The recovered bead coordinates lie within the blood vessels, demonstrating that the Airy-CKM method is correctly identifying the 3D coordinates of the beads. Supplementary \textcolor{blue}{Visualizations 2 \& 3} show multi-view animations of the composite image and beads flowing within SPIM-reconstructed blood vessels.}
\end{figure}

\subsection*{Validation with selective plane illumination microscopy} 
We validated the 3D tracer trajectories obtained with the Airy-CKM technique using Selective-Plane Illumination Microscopy (SPIM), a mature optical-sectioning technique commonly used to image 3D biological samples. SPIM employs a thin sheet of light to illuminate a single plane that is scanned through the volume to yield a 3D image\cite{Huisken:07,Gualda:13}. We built a SPIM illumination arm on top of the sample stage of an inverted microscope (see Methods section for details) to enable quasi-simultaneous optically-sectioned imaging of the zebrafish (Fig. 1(c)). A transgenic zebrafish line expressing green fluorescent protein in the blood vessel walls (flk1:GFP) was used to to facilitate acquisition of the 3D structure of the blood vessels with a $z$ scan subsequent to the Airy-CKM imaging of tracer beads. Validation consisted of assessing whether calculated bead locations lie within the blood volume contained within the vessel walls.

Figure 4 illustrates a superimposition of the 3D tracer trajectories obtained using the Airy-CKM method with a 3D rendering of the $z$-stack data obtained with the SPIM, where a cut-plane visualization in Mayavi is used. As can be observed, the fish was positioned such that the its left-right axis was along the optical axis (i.e., $z$) and its sagittal plane was parallel with the $x$ and $y$ axes. The trajectories lie correctly within the blood vessels, throughout this volume where the intersegmental vessels present a complex three-dimensional structure, indicating that our blood-flow tracking method has good spatial accuracy in both $x-y$ and $z$.

\section*{Discussion}

We have reported high-speed, high-resolution 3D particle tracking in the zebrafish circulatory system by means of pupil-engineered localization microscopy of micro-injected tracer particles. The spatial and temporal resolution of this technique provides an efficient tool for blood-flow characterization that potentially underpins the studies of external mechanical stimuli during the process of vascular formations, and thus will be useful in research into angiogenesis and vasculogenesis.

The major advantage of this technique is the vastly extended detectable volume that can be imaged without significant compromise in precision, which makes it possible to image through thick biological samples without optical sectioning. Snapshot and video-rate 3D particle localization over extended volumes enables imaging of transient phenomena such as, cell adhesion to blood vessel walls, immune cell locomotion and arrhythmic heartbeats.
%
%
%
%

The integration of the cubic-phase mask and lateral beam splitter can be easily implemented with a commercial optical microscope. The 4-$f$ relay can be implemented as a plug-in without modifying the microscope system. In addition, the use of a refractive phase mask provides a much greater optical throughput compared to  diffractive methods, such as a spatial light modulator, while also being much lower cost.

The detectable volume of the Airy-CKM technique is tunable by changing the cubic-phase parameter $\alpha$~\cite{Dowski:95,Zammit:14,Zhou:18}. Larger $\alpha$ yields a larger diffraction-free range in the PSF, allowing a larger operable axial range. For microscopy with a $20\times$, 0.5NA objective lens, the detectable range can be extended to more than \SI{600}{\micro\metre} using a cubic phase stronger than $\alpha=12$ (The conventional depth of field is about \SI{6}{\micro\metre}). However, larger $\alpha$ results in a broader PSF profile and a lower peak SNR, and moreover this is accompanied by a reduced image translation and responsivity to depth. This reduction in image translation can however be compensated by increasing the defocus offset $\Delta\psi$, so that the responsivity to depth can be retained~\cite{Zhou:18}. Overall, the best localization precision is achieved using  the smallest $\alpha$ that ensures a diffraction-free Airy-PSF for the required axial range of localization.

The recovered trajectories include systematic artefacts due to imaging through tissues and organs of differing refractive indices. For example, we observed a predictable and deterministic modulation in the $z$ coordinate of beads flowing in the dorsal aorta (DA) as shown in Fig. 5(a) which may result from the periodic structure of  the ribs and intercostal vessels including the lymphatic vessels. These modulations are not present  in trajectories recorded near the tail of the fish where the tissue macrostructure is more uniform. 
Imaging of the beads within blood containing randomly positioned and oriented blood cells also introduces irregular lensing and phase modulation of the apparent bead positions adding to random uncertainty in the localization accuracy  of about half a micron.

\begin{figure}[ht!]
\centering 
\includegraphics[width=0.57\linewidth]{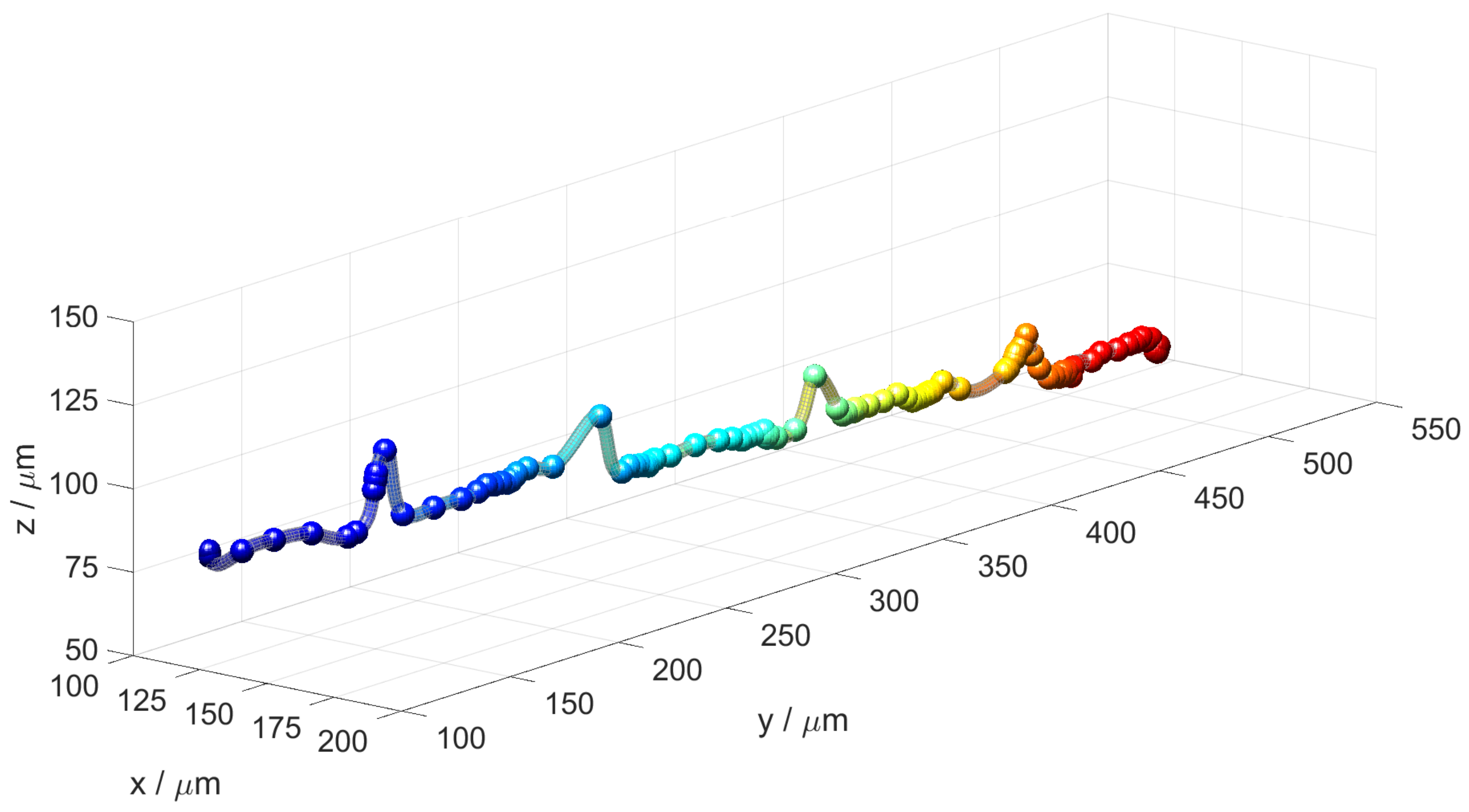}\ \ \includegraphics[width=0.37\linewidth]{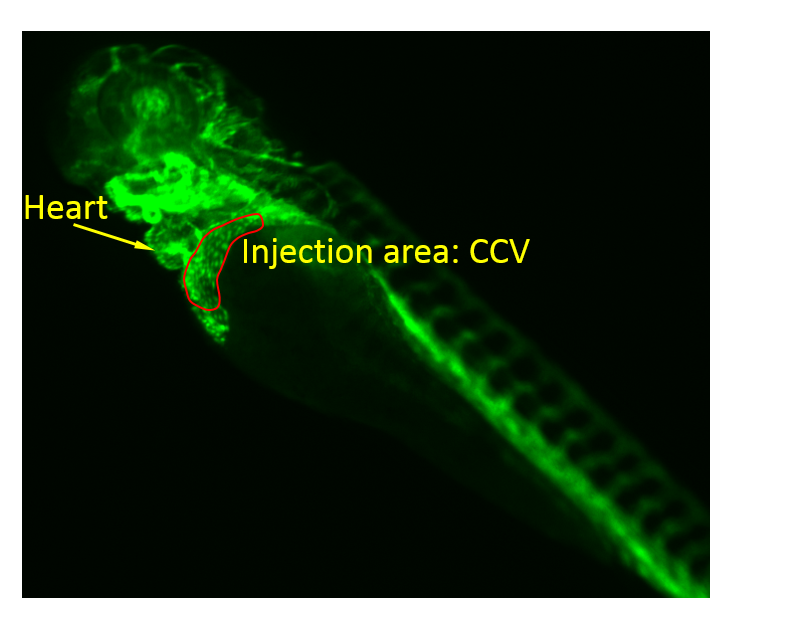}\\ 
\includegraphics[width=0.47\linewidth]{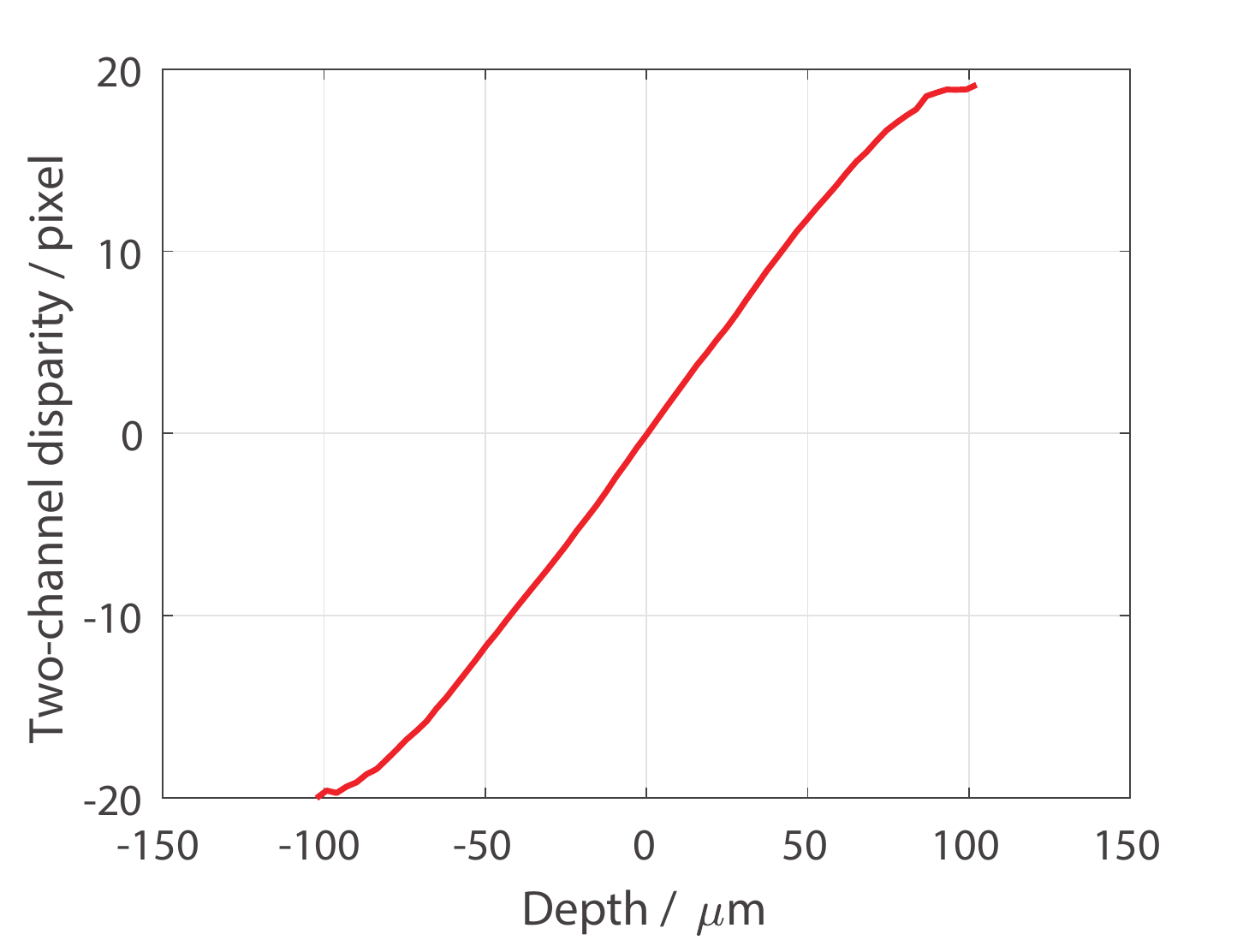}\ \ \includegraphics[width=0.47\linewidth]{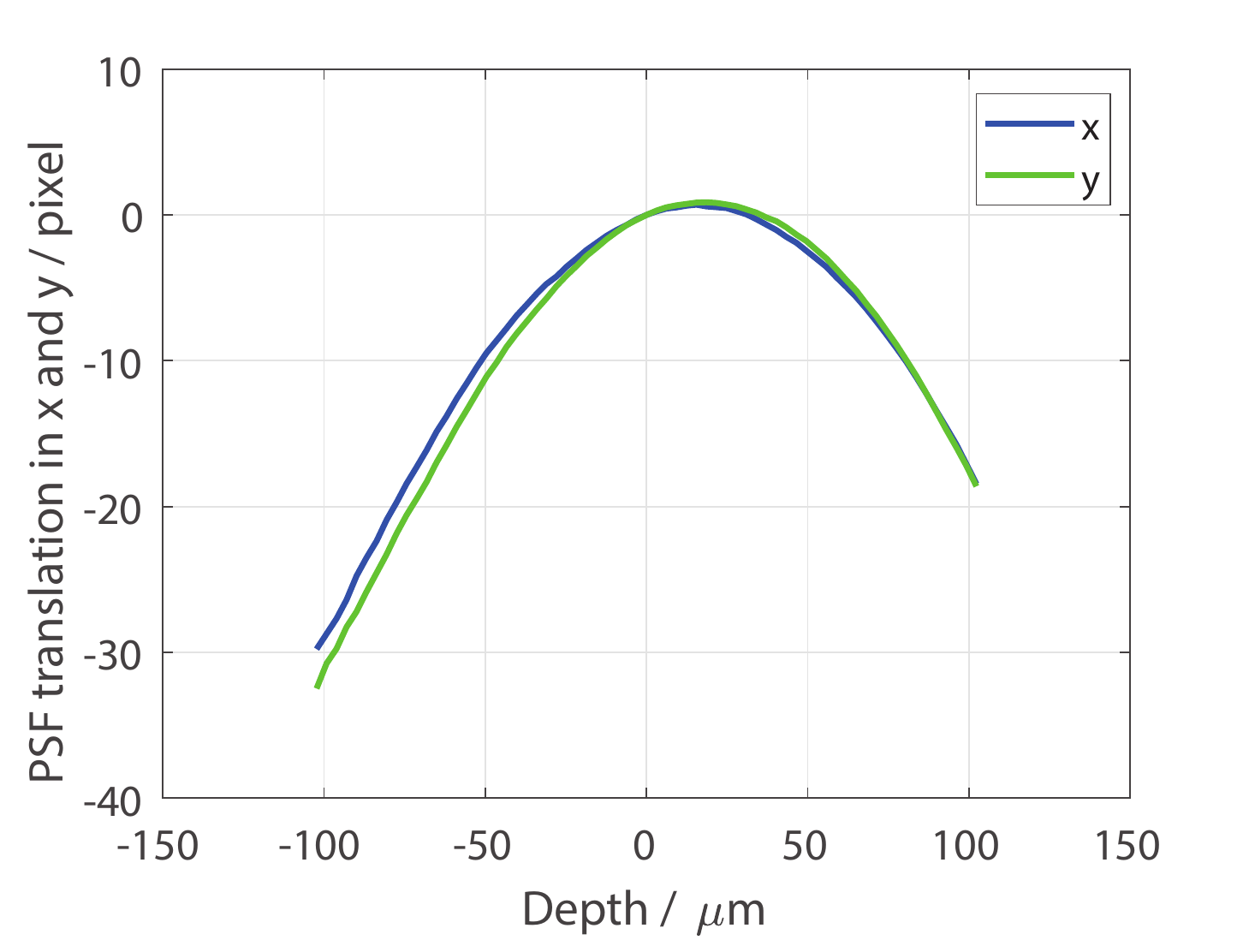}
\put(-170,-10){(c)} \put(-60,-10){(d)} \put(-170,90){(a)} \put(-60,90){(b)}
\caption{\small (a) Trajectory of a bead flowing though the dorsal aorta. Errors in $z$ indicate non-uniformity in the tissue refractive index. (b) Injection area. The micro-injection was performed using a stereo microscope at the common cardinal vein (CCV) indicated with the red curve. Calibration curves of the $20\times$, 0.5NA system with an $\alpha =7$ phase mask. (c) Lookup table relating the two-channel disparity to the $z$ coordinate of the particle. (d) Image translations in $x$ and $y$ as a function of $z$, which are used to correct the image shifts once $z$ is deduced.}
\end{figure}

\section*{Methods}
\subsection*{Micro-injection and sample preparation}

Tracers providing high contrast are crucial to yield precise measurement of the blood flow~\cite{Craig2012}. We employ fluorescent beads as tracers since they are bright and uniform in size and shape, making the particle tracking more robust than the use of endogenous tracers such as the blood cells.

The beads used were \SI{1}{\micro\metre}-diameter fluorescent polymer (Bangslabs, FSEG004 envy green, excited at 525 nm, emission peak at 565 nm), which were selected as their fluorescence emission spectrum is separable from the GFP emission of the zebrafish (flk1:GFP) blood vessel walls (emission peak at 509 nm). The beads were suspended in purified water and washed three times by centrifuging at 9800 relative centrifugal force for 4 minutes to remove additives. Washed beads were resuspended in 1$\times$ phosphate-buffered saline (PBS) solution, at the ratio of 1 part bead solution to 20 parts PBS and sonicated in a water bath sonicator for 10 minutes to prevent bead aggregation before injection.

The prepared solution was then transferred in a glass needle (\SI{10}{\micro\metre}  outer-diameter tip) which was attached to a micro injector (Eppendorf Transjector 5246) with manipulator (Eppendorf Micromanipulator 5171). Three-days-post-fertilization (dpf) transgenic flk1:GFP zebrafish embryos were anesthetized using 200 mg/l Tricaine mesylate solution for 10 minutes before injection (this concentration of Tricaine was maintained throughout the experiment). For a 3 dpf zebrafish, the common cardinal veins (CCV) were chosen as our injection site, since they tend to be flat and thick, lie at the anterior trunk, and are close to the skin of the embryo (Fig. 5(b)). The injection was performed with the assistance of a stereo microscope (Zeiss SteREO Discovery V8).

Note that the injected beads gradually adhere to the blood vessel walls~\cite{Craig2012},  and the time window for observation is therefore limited to about 20 minutes. 

\subsection*{SPIM illumination arm and square capillary mount}
Our SPIM system was modified to work with air objectives for both illumination and detection on a commercial inverted microscope (Nikon Ti Eclipse). The SPIM illumination arm is shown in Fig. 1(c), highlighted with a dashed blue box. A Stradus VersaLase laser system at 488 nm was used as illumination. Two telescopes expand the beam diameter by two and four times respectively, and a 75 mm cylindrical achromat lens (Thorlabs ACY254-075-A) focused the beam in the $y$ direction at the back focal plane of a 10$\times$, 0.3 NA Nikon CFI Plan Fluor DLL objective with a working distance of 16 mm to launch the light sheet. The sheet thickness was measured to be approximately \SI{2}{\micro\metre}. A 20$\times$, 0.5 NA Nikon CFI Plan Fluor objective (the same one used for the Airy-CKM particle tracking) was used to detect the fluorescence signal and an image was then formed on camera 2 with the flipping mirror removed.

The zebrafish embryo was immobilized in 1.5\% agarose after injection and transfered into a hollow square capillary with inner dimensions of 1 mm$\times$1 mm and 0.2 mm wall thickness (VitroCom Inc., 8100-050). The capillary was mounted on a $z$ Piezo stage (PI nano Z Microscope Scanner, P-736.ZR2S) using a custom 3D-printed holder to ensure the walls of the square capillary were well-aligned with the illumination and detection objectives. A $z$ scan of the fluorescent blood vessels was performed by moving the sample in \SI{1}{\micro\metre} steps over a scanning range of \SI{200}{\micro\metre}. The image stack was then processed to reconstruct the vessel structures and to compare with the particle-tracking results.

\subsection*{3D particle localization with Airy-CKM technique}
The excitation for tracking the fluorescent beads was provided by 532-nm diode-pumped solid-state laser (Thorlabs DJ532-40). The beam was diffused with a rotating ground glass and then focused at the back focal plane of the 20$\times$, 0.5NA objective to launch the epi illumination. The fluorescence was collected by the same objective and formed an intermediate image which was then re-imaged by the Airy-CKM particle tracking arm.

The Airy-CKM particle-tracking arm was implemented using a 4-$f$ relay as indicated in Fig. 1(c), using relay lenses with a focal length of 200 mm. A refractive laser-polished cubic-phase mask (custom manufactured by \emph{Power Photonic}, Scotland on a 7 mm$\times$7 mm glass plate) with parameter $\alpha =7$  was placed at the re-imaged pupil plane between two relay lenses to generate the Airy-beam PSF. A lateral beam splitter formed two encoded images with opposite defocus offsets $\psi+\Delta \psi$ and $\psi-\Delta \psi$ respectively. Following Wiener deconvolution the disparity between two recovered images is proportional to the absolute defocus of each point source $\psi$~\cite{Zhou:18}. When an in-focus PSF is used as the deconvolution kernel, the two-channel disparity (TCD) can be written as,
\begin{eqnarray}
TCD\propto\frac{\Delta\psi}{\alpha}\psi.
\end{eqnarray}

\begin{figure}[ht!]
\centering
\includegraphics[width=0.6\linewidth]{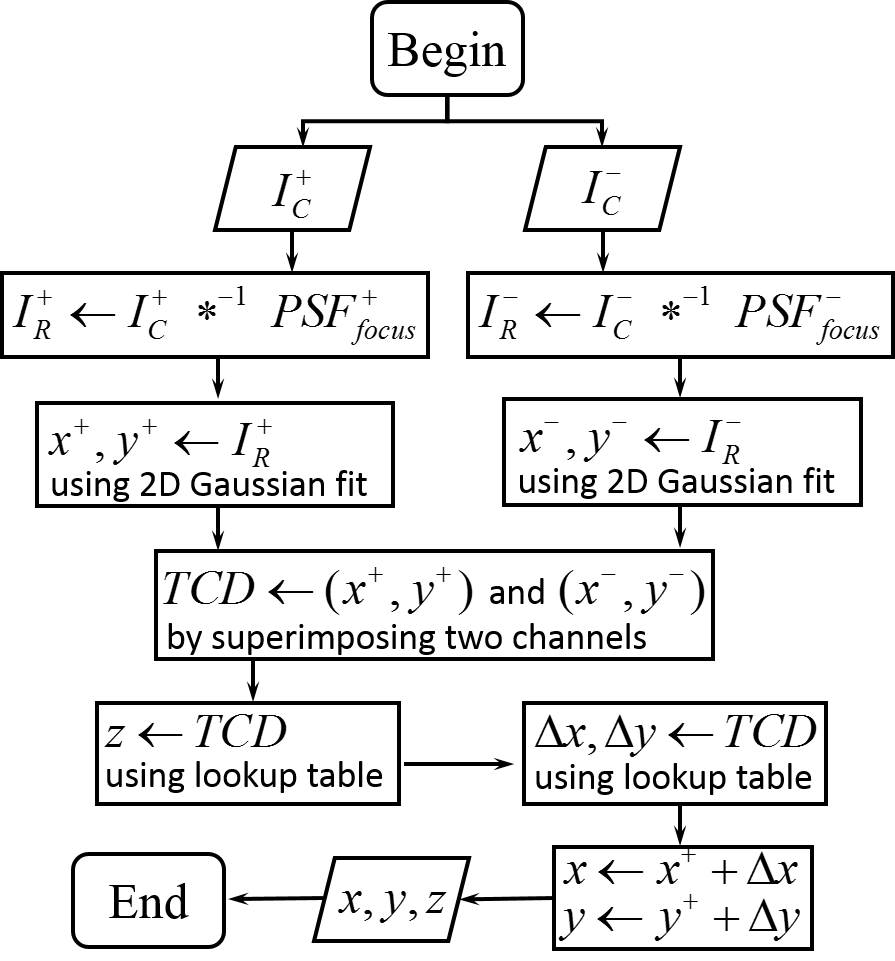}
\caption{\small Algorithm used for 3D particle localization using the Airy-CKM method. $I_{C}^{+}$ and $I_{C}^{-}$ are the recorded images from both channels; $*^{-1}$ refers to the Wiener deconvolution operator; $PSF_{focus}^{+}$ and $PSF_{focus}^{-}$ are PSFs that have been pre-recorded for a fluorescent bead locating at the nominal object plane; $I_{C}^{+}$ and $I_{C}^{-}$ are recovered images by Wiener deconvolution; $\Delta x$ and $\Delta y$ are the shifts in the $x$ and $y$ coordinates due to the PSF translations.}
\end{figure}

Following initial calibration it is possible to calculate the absolute defocus and $z$ coordinate by comparing the $x-y$ coordinates of the tracers in the two imaging channels.
The algorithm used for the image-analysis pipeline is shown in Fig. 6. The recorded images, $I_{C}^{+}$ and $I_{C}^{-}$  from both channels,  are deconvolved with their corresponding in-focus PSFs, $PSF_{focus}^{+}$ and $PSF_{focus}^{-}$, to yield the recovered images $I_{R}^{+}$ and $I_{R}^{-}$. These images approximate a diffraction-limited PSF, as shown in Fig.1 (b) and so 2D Gaussian fit was used to find their $x,y$ coordinates in both channels. The TCD was obtained by superimposing $I_{R}^{+}$ and $I_{R}^{-}$ with the known mapping of the two imaging channels. The TCD together with calibration data (lookup tables were employed) enables the $z$ coordinate and the shifts in  $x-y$ to be deduced, yielding the 3D coordinates.

Figure 5(c) shows the $z$ lookup table (or calibration curve) determined experimentally for the $20\times$, 0.5NA system with $\alpha=7$ phase mask. Since the calibration curve is a monotonic function of depth coordinate, any particle within the \SI{200}{\micro\metre} depth range can be localized without ambiguity. Once $z$ is determined for a particle, its $x$ and $y$ coordinates were corrected to account for the parabolic image translation of the Airy-beam PSF, plotted in Fig. 5(d). The true $x$ and $y$ coordinates are shifted by the amount corresponding to the depth of the particle. The calibration curves were obtained by scanning a single fluorescent bead along the optical axis.

In addition, calculation of the TCD requires the registration of two imaging channels, to calibrate the translational offset and slight difference in magnification between the two channels. This was performed by scanning a fluorescent bead in the nominal object plane throughout the field of view. The recorded images were then deconvolved with $PSF_{focus}^{+}$ and $PSF_{focus}^{-}$ and fit with 2D-Gaussian functions to yield the 2D coordinates  ($x^{+}$,$y^{+}$) and ($x^{-}$,$y^{-}$). An affine function was used to fit the two sets of coordinates, and this function is then considered as the two-channel mapping function.

\section{Funding}
China Scholarship Council (201503170229); British Heart Foundation (NH/14/2/31074); Engineering and Physical Sciences Research Council (EP/M506539/1 and EP/K503058/1).
\section{Acknowledgments}
Animals were maintained according to the Animals (Scientific Procedures) Act 1986, United Kingdom. We are thankful to Charlotte Buckley, Carl Tucker and Martin Denvir for their help and advice with zebrafish breeding.

\bibliographystyle{abbrv}
\bibliography{refs}

\end{document}